\documentclass{epsconf}
\usepackage{graphicx}
\usepackage{epsfig} 
\usepackage{wrapfig}
\usepackage{amsmath}
\title{Dynamics and precise control of fluid V-states using an electron plasma}
\author{\underline{G. Maero}$^{1,2}$, F. Ferrero$^1$, M. Rom\'e$^{1,2}$}
\institute{$^1$ Dipartimento di Fisica `Aldo Pontremoli', Universit\`a degli Studi di Milano, Italy\\
$^2$ INFN Sezione di Milano, Italy}

\begin{document}
\maketitle
Penning-Malmberg traps (PMTs) are electro-magnetostatic devices used to confine nonneutral plasmas at low temperature~\cite{Malm1975}. Their ability to capture and hold bunches of charged particles for a theoretically indefinite time has made them perfect tools to enable fundamental atomic and nuclear physics on matter and antimatter samples. 
Moreover, thanks to the isomorphism between drift-Poisson and Euler equations, inviscid and incompressible 2D fluid experiments can be performed in trapped single-component plasmas: Averaging over the fast axial and cyclotron motion and neglecting inertial effects, in the $\mathbf{E} \times \mathbf{B}$ transverse drift dynamics the density, electric potential and velocity fields of the plasma are analogous to the vorticity, streamfunction and velocity fields of the 2D fluid. This system can actually reproduce an ideal 2D fluid better than laboratory fluid experiments, where 3D and boundary effects are harder to neglect. PMTs also grant the experimentalist a great deal of fine tuning in the parameters of interest~\cite{Dris1990}.

Recent studies have investigated the dynamics and stability properties of vortices undergoing a deformation action by means of static or quasi-static strain flows, obtained applying azimuthally-asymmetric electric potentials to a sectored trap boundary~\cite{Hurs2016,Wong2022}. We intend to analyze the evolution of a vortex where a rotating permanent deformation is generated by inducing a Kelvin-Helmholtz (diocotron) perturbation with a single wavemode. An $l$-fold symmetric uniform vorticity patch in the nonlinear regime, i.e. a generalization of the elliptical Kirchhoff vortex called V-state, was indeed demonstrated to be stable~\cite{Deem1978}, yet a complete understanding of the stability properties, especially depending on the vorticity profile and deformation amplitude, has not been sistematically investigated in controlled experiments so far.

The experiments presented here were performed in the ELTRAP electron plasma trap~\cite{Maer2015}, comprising a stack of $12$ cylindrical electrodes capped on the two ends by a charge collector plate and a phosphor screen, which work as destructive diagnostic tools yielding the charge and the line-integrated density distribution of the confined plasma, respectively, upon release of the trapped sample. Three electrodes are azimuthally split into 2, 4 and 8 sectors, respectively, to be used either as electrostatic signal pickups or to impart non-axisymmetric electric perturbations. The vacuum chamber, at a typical pressure no higher than the mid $10^{-9}$~mbar, is surrounded by a $B\leq 0.2$~T solenoid.

Unlike most electron traps, in our setup the electron plasma is not injected from an external source; the sample is accumulated in situ by ionizing the residual background gas. The combination of ultra-high vacuum and low-amplitude radio-frequency (RF) potential ($1-10$~V, $1-20$~MHz) applied to one of the electrodes in the trapping region makes it possible for free electrons to bounce in the trap for times long enough to be stochastically heated up to energies of some ten electronvolts, allowing them to ionize the residual gas and to accumulate in the trap, while ions are mostly unconfined. Steady-state electron column configurations with densities of some $10^{12}$~m$^{-3}$ and total charge $0.1-1$~nC are obtained in some seconds~\cite{Maer2015}.
\begin{figure}
\centering{\includegraphics[width=0.9\textwidth]{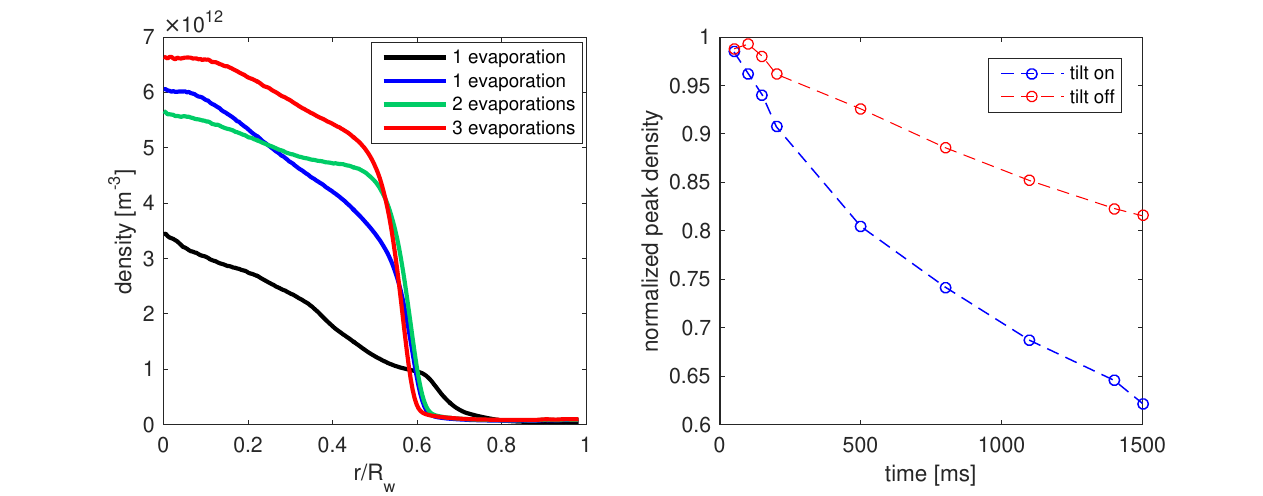}}
\caption{Radial density profile manipulation. Left: Different profiles are achieved playing with the evaporation parameters (residual voltage and ejection time) and number of cycles. Right: The peak density decreases faster when a tilt is induced between trap and magnetic field axis.}
\label{fig:profiles}
\end{figure}

A routine to prepare the initial axisymmetric vortex, i.e. a cold, round column of electrons sitting on the trap's symmetry axis is required as at the interruption of the RF drive the plasma is still hot enough to induce ionization and is often off-axis; furthermore, the presence of residual ions induces a destructive instability of the off-axis bulk rotation~\cite{Paro2014,Maer2017}. Once the RF drive is stopped, the vortex is moved on axis by picking up the rotation signal induced on an electrode sector and sending it to another sector, which progressively reduces the offset. This feedback is maintained for a few seconds to let the plasma cool to a temperature about $1-3$~eV; the poorly confined residual ions in the trapping volume are also lost in this time period. We then shape the radial density profile of the vortex with a temporary, fast reduction of the confining potential and consequently a partial electron ejection or `evaporation'. With one or multiple ejection cycles, quasi-flat or smoothly degrading profiles can be achieved (see left panel of Fig.~\ref{fig:profiles}). A complementary technique under development for the manipulation of the profile is based on the asimmetry-induced radial transport, which can be caused, for instance, by a tilt between the magnetic field lines and the trap axis. In addition to the main solenoid, ELTRAP is equipped with two sets of adjustment coils to correct the tilt; by setting a suitable current in the correction coils for a time of the order of $1$~s, an increased transport can be seen, as shown by the faster decrease of the central density peak in the right panel of Fig.~\ref{fig:profiles}.

The original technique used to excite any $l$-th order Kelvin-Helmholtz (KH) wave is based on the vortex interaction with a rotating electric field. We have shown that a single wavemode can be excited by a field of suitable multipole order rotating with the right orientation (along or against vortex rotation) if the rotation frequency matches the KH wavefrequency. The details of the linear theory and of the experimental implementation have been discussed elsewhere~\cite{Maer2021,Maer2023}. We summarize here some of the most recent and significant results.

The experiments are analyzed with a combination of optical and electrostatic diagnostics. The image of the vortex ejected from the trap is interpolated on a polar grid and Fourier-decomposed in the angular coordinate at each discrete radial position. By formally writing the density map as $n\left( r,\vartheta,t\right) = n^0\left( r\right)+\sum \delta n^l\left( r,t\right)\exp\left[ i\left( l\vartheta-\omega t\right)\right]$, the total deformation amplitude coefficient for any mode of order $l$ is therefore extracted as $A_l\left( r,t\right) = 2\pi\int_{0}^{R_w} \left| \delta n^{l} \left( r,t\right) \right|rdr$.
For modes $l\leq 3$ the use of induced signals on electrode sectors of angular span $\pi$ (dipole configuration) and $\pi/2$ (quadrupole configuration) yields precious information along the whole vortex evolution, and has been used in particular to study the dynamics of the $l=3$ mode. We are working towards the aim of detecting higher-order modes by the implementation of more sensitive signal amplifiers and the installation of a second eight-fold split electrode.

\begin{figure}
\centering{\includegraphics[width=0.9\textwidth]{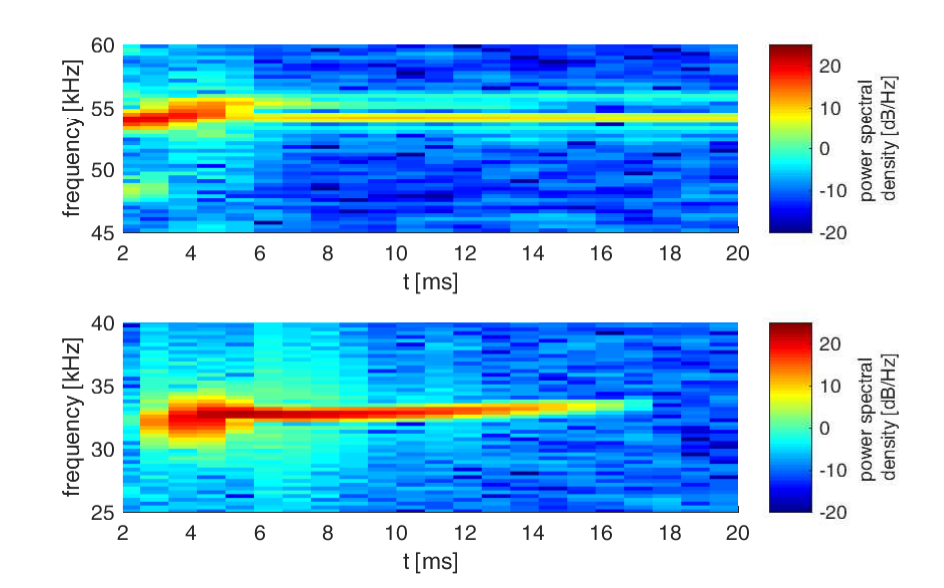}}
\caption{Spectrograms of the electrostatic signals during the rotating field excitation of a mode $l=3$. Upper panel: $l=3$ signal. The persisting line at $54$~kHz is the bleed-through of the drive signal. Bottom panel: $l=2$ signal, induced by the decay of the $l=3$ mother wave.}
\label{fig:spectra}
\end{figure}

The experiments show that in resonance conditions the deformation quickly reaches the nonlinear regime within few tens of rotation periods, even with a weakly perturbing drive (drive to space-charge potential ratio at the vortex edge $\ll 10^{-1}$). The nonlinear regime is characterized by a saturation phenomenon where the tips of the deformation structures (vertices of almost polygonal shapes) start to produce filamentation, which accumulates in a diffuse background and effectively damps the wavemode. At high deformation amplitude, the damping causes catastrophic collapse of the mother $l$-th order wave into the $l-1$-th order daughter wave and a cascade towards axisymmetry is initiated. Also, this regime is evidently nonlinear as the frequency of the parent (and later, daughter) mode is downshifted with increasing amplitude; the spectrograms in Fig.~\ref{fig:spectra} show that the decay of an $l=3$ mode being continuously excited (and reaching saturation approximately in $2$~ms) is indeed accompanied by a rising frequency, as is the $l=2$ mode later on. Preliminary results indicate that the resonance curve of the mode excitation undergoes broadening when the radial vorticity profile is smooth instead of having a sharp edge.

For further studies on the forced and free evolution of V-states, we are aiming for very precise control of the deformation amplitude and to this aim we are implementing a swept-frequency excitation technique where the KH wave can autoresonantly lock to the drive and thus be controlled in amplitude by the drive frequency. This technique has been extensively studied only for the $l=1$ mode~\cite{Faja1999}. Our first results confirm the main features of autoresonance on mode $l=3$, first of all a $3/4$ power law of the drive amplitude threshold value vs the sweep rate. Some peculiar features have also been observed; in particular, as the $l\geq 2$ modes are not bulk mode but edge perturbation that affect the core vorticity when dissipated, we noticed that higher sweep rates make it possible to reach higher deformations; we interpret this as the effect of the competition between mode growth and damping by filamentation.


\begin{thebibliography}{99}
\bibitem{Malm1975}
J.~H. Malmberg and J.~S. deGrassie, Phys. Rev. Lett. {\bf 35}, 577 (1975)
\bibitem{Dris1990}\vspace{-8pt} 
C. F. Driscoll and K.~S. Fine, Phys. Fluids B {\bf 2}, 1359 (1990)
\bibitem{Hurs2016}\vspace{-8pt} 
N.~C. Hurst, J.~R. Danielson, D.~H.~E. Dubin and C.~M. Surko, Phys. Rev. Lett. {\bf 117}, 235001 (2016)
\bibitem{Wong2022}\vspace{-8pt} 
P. Wongwaitayakornkul, J.~R. Danielson, N.~C. Hurst, D.~H.~E. Dubin and C.~M. Surko, Phys. Plasmas {\bf 29}, 052107 (2022)
\bibitem{Deem1978}\vspace{-8pt} 
G.~S. Deem and N.~J. Zabusky, Phys. Rev. Lett. {\bf 40}, 859 (1978)
\bibitem{Maer2015}\vspace{-8pt} 
G. Maero, S. Chen, R. Pozzoli and M. Rom\'e, J. Plasma Phys. {\bf 81}, 495810503 (2015)
\bibitem{Maer2021}\vspace{-8pt} 
G. Maero, N. Panzeri, R. Pozzoli and M. Rom\'e, 47th EPS Conf. on Plasma Phys., ECA {\bf 45A}, P4.4013 (2021)
\bibitem{Maer2023}\vspace{-8pt} 
G. Maero, N. Panzeri, L. Patricelli and M. Rom\'e, submitted to J. Plasma Phys. (2023)
\bibitem{Paro2010}\vspace{-8pt} 
B. Paroli, F. De Luca, G. Maero, R. Pozzoli and M. Rom\'e, Plasma Sources Sci. Tecnol. {\bf 19}, 045013 (2010)
\bibitem{Paro2014}\vspace{-8pt} 
B. Paroli, G. Maero, R. Pozzoli and M. Rom\'e, Phys. Plasmas {\bf 21}, 122102 (2014)
\bibitem{Maer2017}\vspace{-8pt} 
G. Maero, Il Nuovo Cimento C {\bf 40}, 90 (2017)
\bibitem{Faja1999}\vspace{-8pt} 
J. Fajans, E. Gilson and L. Friedland, Phys. Rev. Lett. {\bf 82}, 4444 (1999)
\end{thebibliography}
\end{document}